\documentclass[aps,pra,showpacs,twocolumn,floatfix]{revtex4-1}

\usepackage{epsfig,amsmath,graphicx,float}
\usepackage{amssymb}
\usepackage[normalem]{ulem} 

\usepackage[colorlinks=true,citecolor=blue, linkcolor=blue, urlcolor=blue]{hyperref}

\usepackage[caption=false,position=top,justification=raggedright,singlelinecheck=off,labelformat=parens,subrefformat=parens]{subfig}
\usepackage{color}

\usepackage[usenames,dvipsnames,svgnames,table]{xcolor}
\definecolor{mypurple}{HTML}{b501b5}

\graphicspath{
{./plots/} 
}

\begin{document}

\title{Macroscopic quantum escape of Bose-Einstein condensates: Analysis of experimentally realizable quasi-one-dimensional traps}

\author{Diego A. Alcala$^1$, Gregor Urban$^2$, Matthias Weidem\"{u}ller$^{2,3}$, and Lincoln D. Carr$^{1,2}$}
\affiliation{$^1$Department of Physics, Colorado School of Mines, 1500 Illinois St., Golden, CO 80401, U.S.A.}
\affiliation{$^2$Physikalisches Institut, Universit\"at Heidelberg, Im Neuenheimer Feld 226, D-69120 Heidelberg, Germany}
\affiliation{$^3$Hefei National Laboratory for Physical Sciences at the Microscale, Shanghai Branch, and CAS Center for Excellence and Synergetic Innovation Center in Quantum Information and Quantum Physics, University of Science and Technology of China, Shanghai 201315, China}

\date{\today}

\begin{abstract}

The variational-JWKB method is used to determine experimentally accessible macroscopic quantum tunneling regimes of quasi-bound Bose-Einstein condensates in two quasi one-dimensional trap configurations.
The potentials can be created by magnetic and optical traps; a symmetric trap from two offset Gaussian barriers and a tilt trap from a linear gradient and Gaussian barrier.
Scaling laws in barrier parameters, ranging from inverse polynomial to square root times exponential, are calculated and used to elucidate different dynamical regimes, such as when classical oscillations dominate tunneling rates in the symmetric trap.
The symmetric trap is found to be versatile, with tunneling times at and below one second, able to hold $10^{3}$ to $10^{4}$ atoms, and realizable for atoms ranging from rubidium to lithium, with unadjusted scattering lengths.
The tilt trap produces sub-second tunneling times, is able to hold a few hundred atoms of lighter elements like lithium, and requiring the use of Feshbach resonance to reduce scattering lengths.
To explore a large parameter space, an extended Gaussian variational ansatz is used, which can approximate large traps with Thomas-Fermi profiles.
Nonlinear interactions in the Gross-Pitaevskii equation are shown to produce additional effective mean-field barriers, affecting scaling laws.

\end{abstract}

\pacs{PACS}
\maketitle

\section{Introduction}

Quantum tunneling occurs when an object obeying quantum mechanics penetrates energy barriers which are forbidden by classical analysis, in other words lack sufficient energy to overcome the barriers. 
As a transport phenomenon, this allows movement between wells separated by a classically impenetrable barrier, allowing for recurrence, such as in Josephson junctions. 
Alternatively, it allows decay into free space with a continuum of energies, such as in a quasi-bound state.
Tunneling between finite potentials was originally applied to the study of chiral isomers in a series of papers by Friedrich Hund in 1927~\cite{merzbacher_early_2002}, where he showed that the probability to tunnel through an energy barrier is exponentially dependent on the barrier area.
Alpha decay was understood as a quantum tunneling \emph{escape} process  by  Gamow in 1928~\cite{Gamow1928} and Gurney and Condon in 1929~\cite{Gurney1929}, providing firm evidence that nuclear phenomena were described by quantum mechanics. 
Since its inception, quantum tunneling has found application in several technologies including the scanning tunneling microscope, flash memory, tunneling diodes, and Zener diodes. 
The reader may have noticed that all of these technologies use electron tunneling, the reason being that electrons are very light when compared to atoms.
To be specific, Hund showed that the oscillation frequency of tunneling between two wells (representing left and right-handed isomers) changes from nanoseconds to billions of years with an increase in the barrier area by just a factor of 7.
For this reason, quantum tunneling is considered only relevant for molecules with small energy barriers, such as ammonia which oscillates at $23$ GHz.
Even for single atom tunneling very cold temperatures, very small barriers (tunneling distances), or very light atoms are generally required.
At Kelvin temperatures hydrogen, deuterium, and oxygen tunneling contributes to diffusion in amorphous and polycrystalline ice~\cite{kuwahata_signatures_2015,minissale_quantum_2013}. 
Carbon atoms in complex molecules have been shown to tunnel over sub-angstrom distances at Kelvin temperatures~\cite{zuev_carbon_2003}.
Beyond specific low-temperature systems, quantum tunneling is being experimentally and theoretically shown to affect many organic chemistry and related biological systems, such as in the kinetic isotope effect and enzyme catalysis~\cite{nagel_zachary_d._tunneling_2006,gonzalez-james_experimental_2010,ley_tunnelling_2012}.

Given such apparently stringent constraints on single-particle tunneling, one might hypothesize that quantum tunneling in systems involving larger/heavier atoms, molecules, or many particles with interatomic interactions would be all but negligible. Quite the opposite, tunneling in such systems, termed macroscopic quantum tunneling (MQT), encompasses a much richer landscape due, but not limited, to the statistics of the particles (whether fermionic, bosonic, or anyonic), larger masses leading to gravitational effects, weak interatomic interactions allowing quasi-particle descriptions (magnons, polaritons, excitons, etc), strong interactions dominated by many-body effects (fluctuations, entanglement, strongly correlated systems, etc), and the interaction between particles and time-dependent potentials, for a detailed overview see~\cite{zhao_macroscopic_2017}.
We will focus on studying the regime of repulsive weakly-interacting bosonic atoms, specifically in a dilute ultracold gas. Such a system can be adequately treated via a semi-classical field which assumes quantum fluctuations around the mean are negligible: a mean-field description.

Ultracold atomic systems are ideal for exploring the MQT landscape, with experimental access to single-atom systems, such as with optical tweezers, up to billions of atoms.
Josephson tunneling, where two weakly linked macroscopic wave functions undergo MQT, has been measured for both AC and DC configurations in superconductors~\cite{likharev_superconducting_1979} and superfluid He~\cite{pereverzev_quantum_1997,sukhatme_observation_2001}.
The Josephson effect has been used to create superconducting quantum interference devices, to measure the value of the volt, as a measure of elementary charge, and is important for many potential quantum computing applications~\cite{makhlin_quantum-state_2001}.
A Bose-Einstein condensate (BEC) is an object composed of $10^{3}$ to $10^{9}$ atoms~\cite{greytak_boseeinstein_2000}, with most in the same quantum mechanical state; in other words collectively behaving as a single large quantum object.
Many properties make BECs particularly suited to study MQT. Highly controllable atom-atom interactions can be tuned over several orders of magnitude via Feshbach resonances~\cite{schweigler_experimental_2017}.
Advances in optical, magnetic, and radio-frequency traps allow for a diversity of trap geometries and highly controllable experiments~\cite{fernholz_dynamically_2007,ashkin_observation_1986,bloch_many-body_2008}.
Time-of-flight measurements are able to measure first and second order correlations~\cite{bloch_many-body_2008}, important to distinguish many-body effects. Interference patterns have been used to measure up to 10th order correlations in superfluids~\cite{schweigler_experimental_2017}.
MQT in \emph{double-well} systems is well established in BECs for both the AC and DC Josephson effects~\cite{albiez_direct_2005,levy_.c._2007}, with interactions allowing for self-trapping regimes and decreased  oscillation period by an order of magnitude.
Furthermore, the first mean-field or semi-classical observation of \emph{quantum tunneling escape} has been made~\cite{potnis_interaction-assisted_2017}, where interactions produced non-exponential decay.

In this Article we will explore macroscopic quantum escape in two quasi-1D trapping configurations: a \emph{symmetric trap} using offset Gaussians and a \emph{tilt trap} using a Gaussian and linear gradient. Both of these are experimentally realizable using a combination of magnetic and optical traps.
While there are many techniques to go beyond mean-field physics, such as density-matrix renormalization group~\cite{white_density_1992}, multiconfigurational time-dependent Hartree methods~\cite{lode_how_2012,beinke_many-body_2015}, Quantum Monte Carlo~\cite{prokofev_exact_1998,sandvik_quantum_1991}, and dynamical mean field theory~\cite{georges_dynamical_1996}, we choose a combination of the variational principle and a modified Jeffreys-Wentzel-Kramers-Brillouin (JWKB) model (variational-JWKB), which includes mean-field effects, to understand the gross features of MQT out of a trapping potential. This variational-JWKB method~\cite{moiseyev_transition_2004,carr_macroscopic_2005} allows for rapid exploration of large parameter spaces, compared to more powerful and expensive numerical techniques which explicitly account for the many-body effects.
Moreover, most BEC experiments are sufficiently dilute so that corrections to the mean-field energy and chemical potential are less than 1\%, and depletion due to correlations are also less than 1\%; contrast this to superfluid $^{4}\mathrm{He}$, where depletion is near 90\%. 
Whether mean-field theory accurately describes MQT dynamics, or many-body effects dominate, is an open question, but one cannot answer this clearly without first having a thorough picture of mean field effects.
We find scaling laws using experimentally controllable trapping parameters, obtaining regions where scaling in trap parameters is dominated by the classical oscillation period in the trap or tunneling probability through a barrier, and determine accessible experimental conditions for MQT realization.
Furthermore, we find mean-field interactions cause the appearance of additional effective barriers within the trap for certain parameter ranges, altering the overall tunneling rates and scaling laws.

This Article is organized as follows. First, we introduce the variational-JWKB formalism, discuss the barrier configurations under study, and outline our numerical procedure in Sec.~\ref{sec:Variational JWKB}. With the formalism and numerics fleshed out, we present scaling laws and discussions on implementation for the symmetric and tilt potentials in Sec.~\ref{sec:Scaling Symmetric}. 
Afterwards, considerations and findings from the variational-JWKB method -- effects of different variational functions and the appearance of additional wells in the effective potential -- are discussed in Sec.~\ref{sec:Variational JWKB Considerations}.
Finally, we summarize our findings in Sec.~\ref{sec:Conclusion}.

\section{Variational JWKB Method}
\label{sec:Variational JWKB}

In this section, we overview the variational-JWKB method employed and present the two barrier configurations under study, along with the wave functions used in all figures; we also explore other ans\"atze in Sec.~\ref{sec:Considerations_WF}.
We illustrate the  numerical procedure in detail for the tilt potential.

\subsection{Formalism}

The variational-JWKB method consists of two procedures, variationally finding a wave function for the Gross-Pitaevskii equation (GPE), Eq.~\eqref{eq:gpe}, and using a modified JWKB method to calculate the tunneling rate, i.e. the complex component of the chemical potential. A dilute and weakly interacting Bose gas at zero temperature, ignoring quantum fluctuations~\cite{dalfovo_theory_1999}, is described by
\begin{equation}
\left(-\frac{\hbar^2}{2m} \frac{\partial^2}{\partial{x}^2} + V(x) + {g}\vert\psi(x)\vert^2\right)\psi(x)=\mu\psi(x) \label{eq:gpe},
\end{equation}
a quasi-1D time-independent GPE, with tight harmonic confinement assumed in the transverse direction~\cite{carretero-gonzalez_nonlinear_2008}, 
under the assumptions of separation of variables in time and space, where the chemical potential $\mu$ is taken to be complex to capture quasibound or decaying states undergoing MQT escape.
The single-particle wave function is given by the order parameter, $\psi$, and the total number of atoms, $N$, by normalization $\int |\psi(x)|^{2} dx = N$.
Assuming binary contact interaction between atoms gives the nonlinear interaction parameter, $g$. This is related to the full 3D parameter via $g = g_{\mathrm{3D}}/2 \pi \ell_{\bot}^{2}$, with $g_{\mathrm{3D}}=4 \pi \hbar^{2}  a_{\mathrm{s}}/m$, transverse harmonic oscillator length $\ell_{\bot} = \sqrt{\hbar/m \omega_{\bot}}$, $\hbar$ the reduced Plank constant, $m$ the atomic mass, $\omega_{\bot}$ the transverse confining angular frequency, and $a_{\mathrm{s}}$ the $s$-wave scattering length. The external 1D confining potential from which the atoms will tunnel, either in the symmetric or tilt configuration, is given by $V(x)$.

The first step in our calculation is to find a metastable state for Eq.~\eqref{eq:gamma}, using a variational wave function, $\psi(x;\alpha_1,...,\alpha_M)$, with variational parameters $(\alpha_1,...,\alpha_M)$ chosen under consideration of the potential, presented in Sec.~\ref{sec:bar_config}.
There exists a caveat in using the variational procedure for this nonlinear problem, specifically that the wave function is not normalized until after variation~\cite{malomed2002variational}. Using the non-normalized wave function, the total system energy, $E(\alpha_1,...,\alpha_M)$, is calculated, 
\begin{align}
&E(\alpha_1,...,\alpha_M) = \int\limits_{-\infty}^{\infty}{\mathcal{E} dx} , \label{eq:energy_integral} \\
&\mathcal{E}=\frac{\hbar^2}{2m}\vert\nabla\Psi(x)\vert^2 + V(x)\vert\Psi(x)\vert^2 + \frac{1}{2}g\vert\Psi(x)\vert^4 , \label{eq:gpe_energy}
\end{align}
assuming $\lim_{x \to \pm \infty} \Psi(x) = 0$.
Varying the energy with respect to $(\alpha_1,...,\alpha_M)$ and calculating the normalization condition produces a system of $M+1$ equations, the solution procedure for this system is fully discussed in Sec.~\ref{sec:num_procedure}.

After a variational solution, $\psi(x)$, has been found, the tunneling rate is calculated using a modified JWKB procedure, accounting for the mean-field interaction. The standard JWKB tunneling rate, $\Gamma= ($Average barrier collisions frequency$) \times ($Tunneling probability$)$, is given by,
\begin{align}
\Gamma &= \nu  \exp \biggl(-\frac{2}{\hbar} \int\limits_{x_\mathrm{in}}^{x_\mathrm{out}} \mathrm{dx} |p(x)| \biggr), \label{eq:gamma}\\
\nu &= \biggl(m \oint \frac{\mathrm{dx}}{|p(x)|} \biggr)^{-1}, \label{eq:oscillation}
\end{align}
where higher order corrections in the actions are negligible for our purposes.

In Eq.~\eqref{eq:gamma}, $x_{\mathrm{in}}$ and $x_{\mathrm{out}}$ are the classical turning points at the inner and outer edges of the barrier, found by solving for zero-momentum, $p(x_\mathrm{in})=p(x_\mathrm{out})=0$. Equation~\eqref{eq:oscillation} is the semi-classical oscillation frequency in the well, the momentum, $p(x)$, is given by the standard form as a function of total (in this case chemical) and potential energy, except the potential $V(x)$ is replaced with the effective potential $V_{\mathrm{eff}}(x)$,
\begin{align}
p(x) &= \sqrt{2m \left( V_{\mathrm{eff}}(x) - \mu \right)}, \label{eq:momentum}\\
V_{\mathrm{eff}}(x) &= V(x) + g{\left|\Psi(x)\right |}^2 \label{eq:eff_pot},
\end{align}
accounting for inter-particle interactions.

We assume that, on average, the atoms move with the semi-classical momentum $p(x)$ in the effective potential $V_{\mathrm{eff}}(x)$, a good approximation for BECs, which are highly condensed into a single energy state. The average period between collisions with the outermost barrier, the barrier from which the particles can escape, is $\nu^{-1}$.  The closed path in Eq.~\eqref{eq:oscillation} indicates that the integral is calculated over a period of oscillation between barriers from which the atoms can escape. For the symmetric barrier, one period of oscillation would be between the barriers, i.e., from the left barrier to the right one or vice-versa, while for the tilted barrier one period of oscillation would be from the right barrier, to the left barrier, and back to the right barrier, Fig.~\ref{fig::TiltPot}. Using this effective mean-field barrier produces a surprising result.
It can give rise to additional effective barriers, Fig.~\ref{fig::plot_mean-field-barrier}, which have a noticeable effect on tunneling rates and require a slight modification of Eq.~\eqref{eq:gamma}, as detailed in Sec.~\ref{sec:eff_mult_wells}.

\subsection{Barrier Configurations and Ansatz}
\label{sec:bar_config}

\begin{figure}
	\centering
	\includegraphics[width=1.0\linewidth]{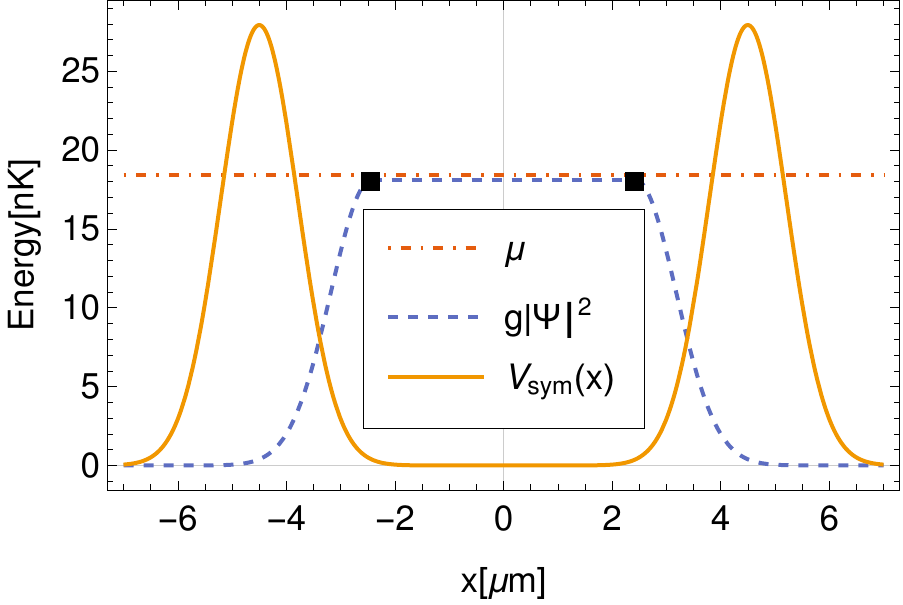}
	\caption{\label{fig::SymPot}
		\emph{Symmetric trap variational wave function.}
		A typical wave function solution, $g |\Psi|^{2}$ (dashed blue line), and chemical potential, $\mu$ (dash-dotted red line), for $V_{\mathrm{sym}}(x)$ (solid yellow line) with parameters $V_{0} = 27.9 \mathrm{[nK]}$,  $\sigma_{0} = 1.41 \mathrm{[\mu m]}$, and $x_{0} = 4.5 \mathrm{[\mu m]}$ are shown. Black squares mark the connection points in the wave function between the two Gaussian tails for $|x| > 2.43$ and the flat region in the middle.
	}
\end{figure}

We investigate two potential configurations which lend themselves well to experimental implementation. First, a \emph{symmetric} well, $V_{\mathrm{sym}}$, is formed by two displaced Gaussian barriers, given by Eq.~\eqref{eq:SymPot}. This models, for example, a 1D slice through a rotationally symmetric trap, or a trap formed by two light sheets~\cite{potnis_interaction-assisted_2017}. The second \emph{tilt} configuration, $V_{\mathrm{tilt}}$, is created from a Gaussian barrier with a linear ramp given by Eq.~\eqref{eq:TiltPot}, where the tilt could result from gravity and/or a magnetic field,
\begin{align}
V_{\mathrm{sym}}(x) & =  V_0 \left[ e^{-(x-{x_0})^2/{2 \sigma_{0}}^2} +  e^{-(x+{x_0})^2/{2 \sigma_{0}}^2} \right] \label{eq:SymPot} \\
V_{\mathrm{tilt}}(x) & =  V_0 e^{-x^2/{2 \sigma_{0}}^2} - \alpha_{0} x \label{eq:TiltPot}.
\end{align}

The wave function ansatz for the symmetric well is given by Eq.~\eqref{eq:Gauss_WF}, with variational parameters being the Gaussian separation $x_{1}$, width $\sigma_{1}$, and amplitude $A$. In other words, variation in the energy density, Eq.~\eqref{eq:gpe_energy}, is performed with respect to $A$, $x_{0}$, and $\sigma_{0}$; a typical solution is given in  Fig.~\ref{fig::SymPot}. Physical motivations for this ansatz split the wave function into three regions: a flat middle section, and left/right decaying wave function tails, which extend through the trapping barriers. The flat middle section mimics the Thomas-Fermi approximation when the barrier separation is large~\cite{bloch_many-body_2008}, but also allows for a nearly Gaussian solution for narrow traps; details of how a pure Gaussian ansatz underestimates escape rates in a wide well are presented in Sec.~\ref{sec:Considerations_WF}.
\begin{equation}
\Psi(x)= 
\begin{cases}
A\; \exp\left({-|x - x_{1}|^2/2\sigma_{1}^2}\right)&  |x|> x_{1}\\
A              & |x| \leq x_{1} \\
\end{cases} \label{eq:Gauss_WF}
\end{equation}
Beyond the commonly used Gaussian tails ansatz, we also explored modifying these tails using a superposition of exponential and Gaussian functions, Sec.~\ref{sec:Considerations_WF}.

\begin{figure}
	\centering
	\includegraphics[width=1.0\linewidth]{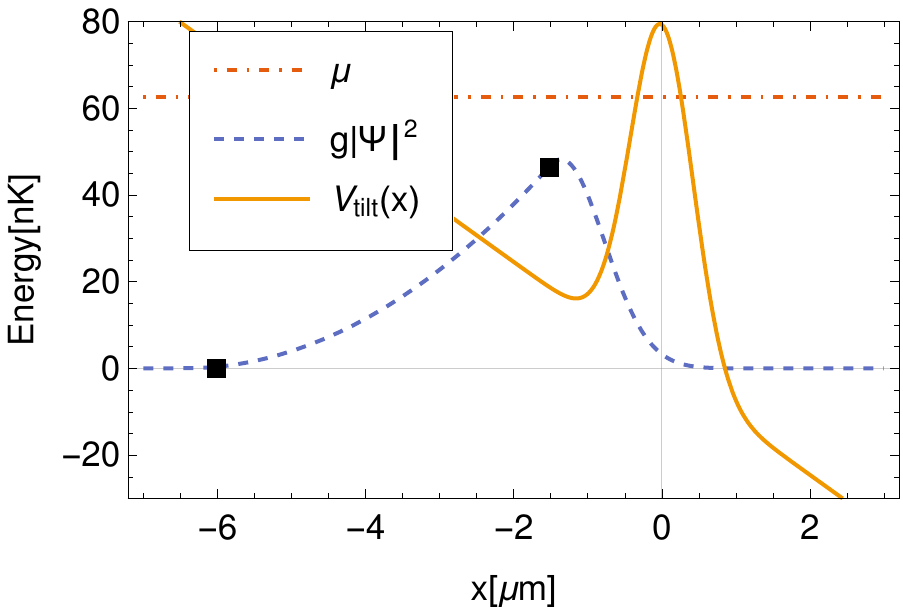}
	\caption{	\label{fig::TiltPot}
	\emph{Tilt trap variational wave function.}
	A typical wave function solution, $g |\Psi|^{2}$ (dashed blue line), and chemical potential, $\mu$ (dash-dotted red line), for $V_{\mathrm{tilt}}(x)$ (solid yellow line) with parameters $V_{0} = 79.3 \mathrm{[nK]}$, $\sigma_{0} = 0.85 \mathrm{[\mu m]}$, and $\alpha_{0} = 1.07 \mathrm{[m/s^{2}]}$ are shown. Black squares show the connection points for the different wave function pieces; the left Gaussian tail below $x \approx -6$, middle connection region between $x \approx -6$ and $x \approx -1.5$, and right Gaussian tail above $x \approx -1.5$.}
\end{figure}

The tilt barrier ansatz is given by Eq.~\eqref{eq:Tilt_WF}. Similar to the symmetric potential, we allow a middle \emph{linear} region which can mimic the Thomas-Fermi approximation for wide traps, but allow the middle region to have a nonzero slope since the barrier is not symmetric; $V_{\mathrm{tilt}}(x)$ is thus formed by two independent Gaussian tails, connected by a linear function,
\begin{equation}
\Psi(x)= 
\begin{cases}
B_{L}\; \exp\left({-(x + x_{L})^2/2 \sigma_{L}^2}\right)&  x< C_{L}\\
A_{1} x + A_{2},              & C_{L} \leq x \leq C_{R} \\
B_{R}\; \exp\left({-(x - x_{R})^2/2 \sigma_{R}^2}\right)&  x> C_{R}. \\
\end{cases} \label{eq:Tilt_WF}
\end{equation}
A typical solution for $V_{\mathrm{tilt}}$ is shown in Fig.~\ref{fig::TiltPot}. The variational parameters are the Gaussian heights for the left and right tails, $B_{L}$ and $B_{R}$, displacement of the tails, $x_{L}$ and $x_{R}$, width of the tails, $\sigma_{L}$ and $\sigma_{R}$, and the slope $A_{1}$ and intercept $A_{2}$ of the middle region. Continuity conditions on the wave function at the connection points, $C_{L}$ and $C_{R}$, will eliminate some of the previously stated parameters and result in $C_{L}$ and $C_{R}$ becoming variational parameters.

\subsection{Numerical procedure}
\label{sec:num_procedure}

We illustrate the variational-JWKB procedure, varying the wave function ansatz to find a metastable solution and applying the modified JWKB procedure, for $V_{\mathrm{tilt}}$.
The unknowns are the wave function parameters ($A_{1}, A_{2}, B_{L}, B_{R}, x_L, x_R, \sigma_L, \sigma_R$), the connection points ($C_{L}$ and $C_{R}$),  and the total energy. Trap parameters and nonlinearity ($V_0, \sigma_{0}, \alpha_{0}, g$) are input parameters, chosen under considerations of the experiment.

The variational component involves deriving and solving a system of equations. First we consider wave function boundary conditions; match the height and derivative of the wave function at the two connecting points ($C_{L}$ and $C_{R}$). The four boundary conditions allow us to solve for four parameters; we choose $A_1$  $A_2$, $B_{R} $, and $x_{R}$, noting that the connecting points now become variational parameters in the wave function. 
Boundary conditions can be used to eliminate any variable, but these were chosen for algebraic simplicity.
Second, we calculate the total energy of the wave function, Eq.~\eqref{eq:energy_integral}. 
Third, we calculate the first and second derivatives of the energy with respect to remaining unknown wave function parameters $( B_{L}, x_L, \sigma_{L}, \sigma_{R}, C_{L}, C_{R})$, along with the Hessian.
From this procedure, we now have seven equations (first derivatives must be zero, and normalization condition), seven unknown parameters $( B_{L}, x_L, \sigma_{L}, \sigma_{R}, C_{L}, C_{R}, \mu)$, and eight constraints (second derivatives and Hessian are positive for minimum energy). We emphasize that the normalization condition must not be enforced until after variation of the energy. After variation, we can apply the normalization condition, resulting in unknowns  $( \mu, x_L, \sigma_{L}, \sigma_{R}, C_{L}, C_{R})$, which can be numerically found.

Given trap parameter values ($V_0, \sigma_{0}, \alpha_{0}$) and the non-linearity, $g$, solutions for all wave function parameters are found numerically to hold (at least) to $10^{-60}$ absolute accuracy (the working precision is chosen high enough) and checked to fulfill the constraints. With numerical solutions to all parameters, we proceed with the modified JWKB calculation, Eq.~\eqref{eq:gamma}. Because we use the effective potential in the momentum, Eq.~\eqref{eq:momentum}, additional effective mean-field barrier \emph{islands} inside the potential appear in the calculations, Sec.~\ref{sec:eff_mult_wells}.

This procedure can then be repeated for all desired potential parameters and interactions. As the minimization problem is not convex, a solution-caching system was used. Variational parameters to a trap configuration are found using the solution of a close trap configuration as a starting point of the search; the first solutions used in this system were thoroughly checked to be minimized metastable states, and the distance between adjacent parameter values ($V_0, \sigma_{0}, \alpha_{0}, g$) was kept small. For most parameter scans, barrier parameters were held fixed while the nonlinearity was incremented in small steps relative to the cutoff nonlinearity, where the repulsive mean-field interaction become too large for trapped states.

\section{Barrier Analysis}
\label{sec:Scaling Symmetric}

In this section we explore the parameter space of the two barrier configurations given by the experimentally adjustable Gaussian width, $\sigma_{0}$, Gaussian height, $V_{0}$, mean-field interaction, $g$, and separation $x_{0}$ (acceleration $\alpha_{0}$) for $V_{\mathrm{sym}}$ ($V_{\mathrm{tilt}}$).
We present scaling laws for the maximal tunneling rate in $x_{0}$, $V_{0}$, and $\alpha_{0}$. Furthermore, we analyze experimental implementation and limitations of the potentials, presenting the parameter regimes that lend themselves well to studying MQT. All plots here are for rubidium; we analyze how these change for lithium and sodium in Sec.~\ref{sec:Experimental_Conditions}.

\subsection{Scaling laws for Symmetric Barrier}

\begin{figure}
	\subfloat{
		\includegraphics[width=1.0\linewidth]{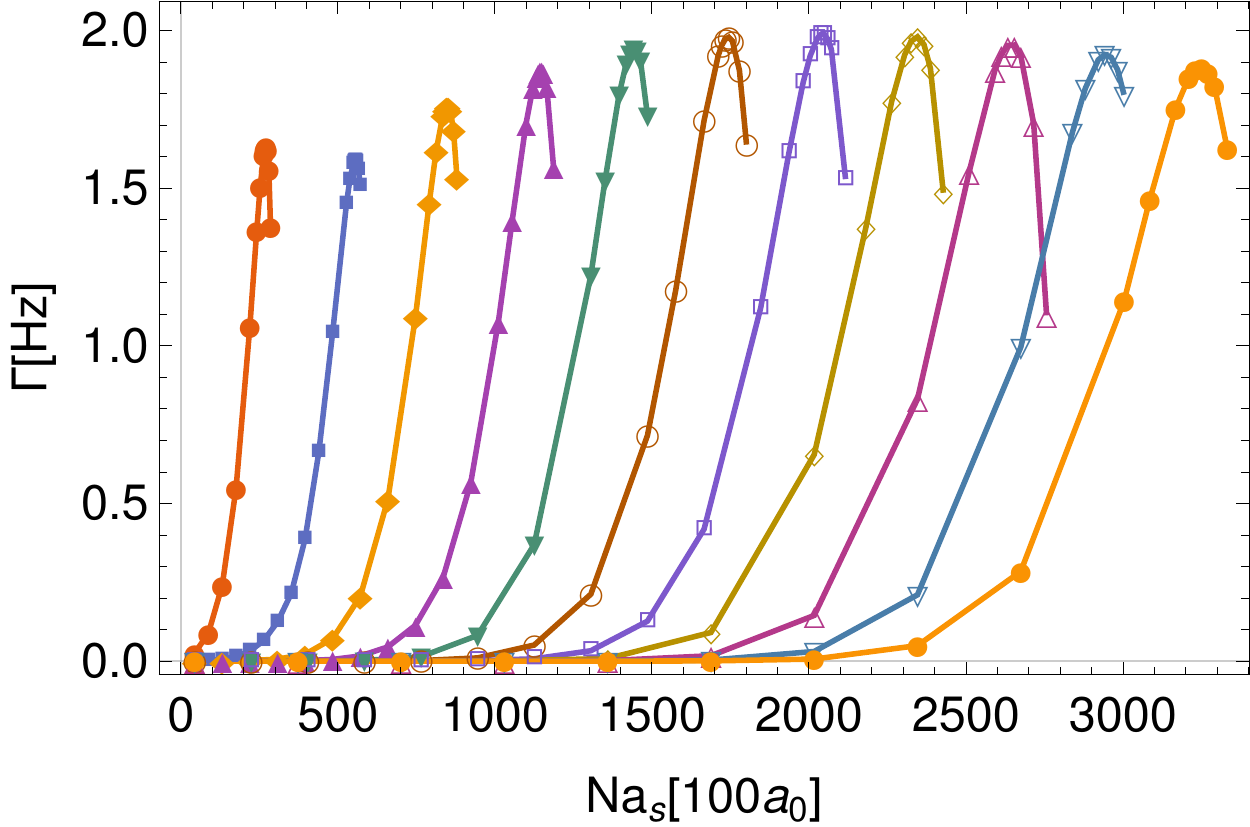}
	}
	\caption{\label{fig::gamma_vs_g_various}
		\textit{Tunneling rate for various barrier heights.}  Tunneling rate vs interaction strength, given by number of atoms $N$ and scattering length $a_{s}$ in units of Bohr radius $a_{0}$, is shown for $V_{\mathrm{sym}}$ with 11 equally spaced barrier heights between $V_{0} = 14.0 $ and $ 125.6 \mathrm{[nK]}$ with parameters $x_{0}=4.5 \mathrm{[\mu m]}$ and $\sigma_{0} = 1.41 \mathrm{[\mu m]}$, assuming transverse frequency $\omega_{\bot}=2 \pi \times 500 \mathrm{[Hz]}$. Larger barriers allow for greater mean-field interaction. The maximal tunneling rate shows a non-trivial trend, initially decreasing, followed by increasing, and then decreasing again. For rubidium, the largest barrier $V_{0} = 125.6 \mathrm{[nK]}$ would allow $N \approx 3000$. Markers are data with error bars smaller than marker, and curves are a guide to the eye.
	}
\end{figure}

\begin{figure}
	\subfloat[]{
		\includegraphics[width=1.0\linewidth]{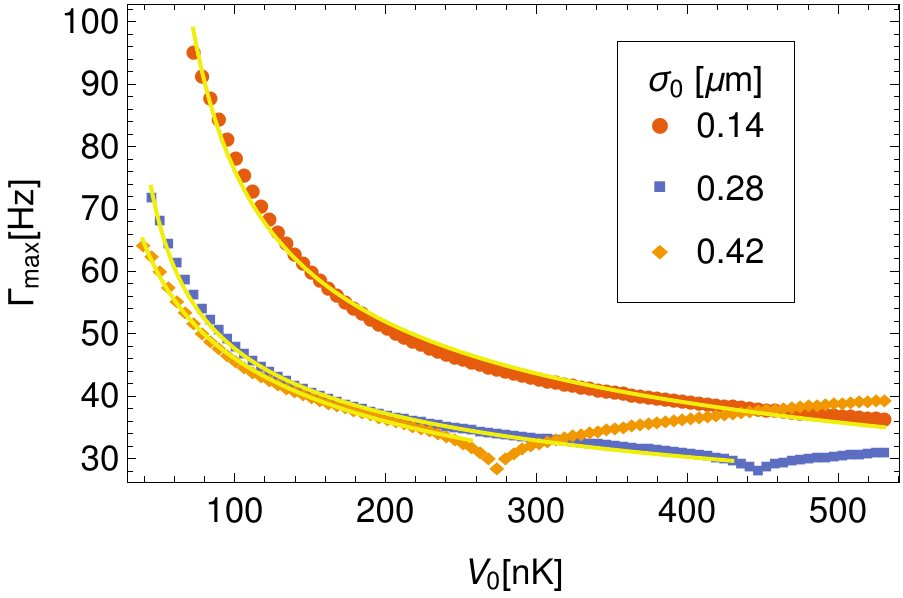}
	}
	\hfill
	\subfloat[]{
		\includegraphics[width=1.0\linewidth]{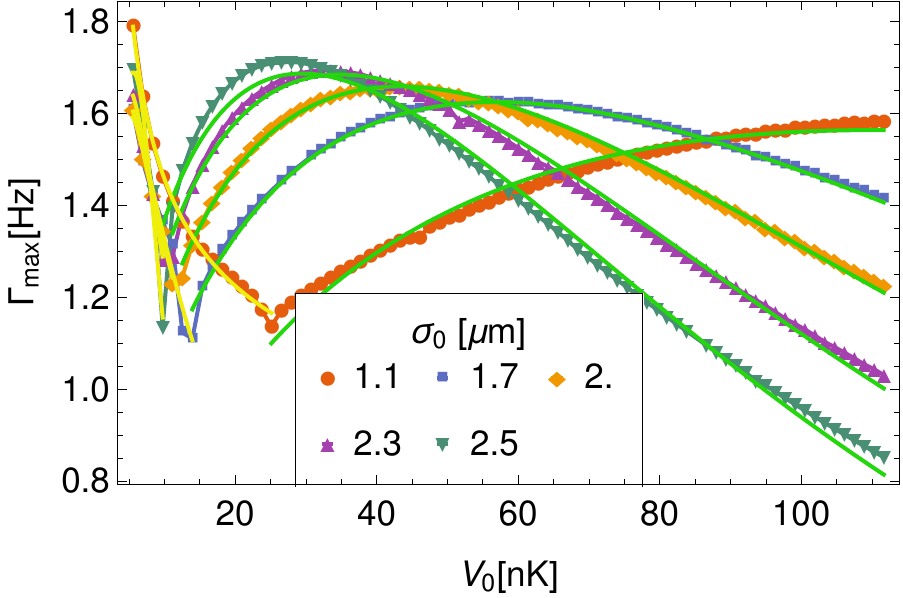}
	}
	\caption{\label{fig::sympot_v1_scaling}
		\textit{Scaling in barrier height for symmetric trap.} Maximal tunneling rate as a function of $V_{0}$ for different values of $\sigma_{0}$, with barrier separation for (a) $x_{0}=1.0\mathrm{[\mu m]}$ and (b) $x_{0}=5.0\mathrm{[\mu m]}$; solid yellow (green) curves are fits before (after) kink, emergence of effective mean-field islands. For smaller barrier width $x_{0}=1.0 \mathrm{[\mu]}$, scaling is dominated by the classical oscillation frequency. Error bars for data points are smaller than markers.
	}
\end{figure}

Adjustable parameters, those that can be controlled experimentally, for the symmetric barrier are the trap parameters ($V_{0},x_{0},\sigma_{0}$) and the interaction strength, $Ng$, in units of scattering length $a_{s}$, $Ng = 2 \hbar \omega_{\bot} N a_{s}$. 
We first look at how tunneling rate, $\Gamma$, depends on barrier height, $V_{0}$, and mean-field interaction, $N g$, in Fig.~\ref{fig::gamma_vs_g_various} for $x_{0}=4.5\mathrm{[\mu m]}$ and $\sigma_{0}=1.41 \mathrm{[\mu m]}$. All barrier heights have an interaction strength, $g_{\mathrm{max}}$, beyond which bound modes are no longer supported, with larger barriers having larger overall $g_{\mathrm{max}}$; this maximal value occurs because we are considering repulsive interactions, $g>0$, which can become strong enough to overcome the trapping barrier. With increasing $V_{0}$, one might expect that tunneling rates will generally be smaller; larger barriers are more difficult to penetrate.
Maximal tunneling rates, $\Gamma_{\mathrm{max}}$,  generally decrease with increasing $V_{0}$ in Fig.~\ref{fig::sympot_v1_scaling}, except for a barrier range where $\Gamma_{\mathrm{max}}$ increases with increasing $V_{0}$.
This phenomenon is caused by the appearance of additional mean-field barrier \emph{islands}, turning the single well into a multiple well, Fig.~\ref{fig::plot_mean-field-barrier}; this requires a slight modification in the calculations, as fully discussed in Sec.~\ref{sec:eff_mult_wells}. We find similar trends when examining $\Gamma$ vs $g$ for different $x_{0}$.
Tunneling rates for any given parameter configuration are strongly peaked around some mean-field strength, $g_{0}$. For this reason, we find scaling laws in $\Gamma_{\mathrm{max}}$ which can be used to seek appropriate MQT regimes.

\begin{figure}
	\subfloat[]{
		\includegraphics[width=1.0\linewidth]{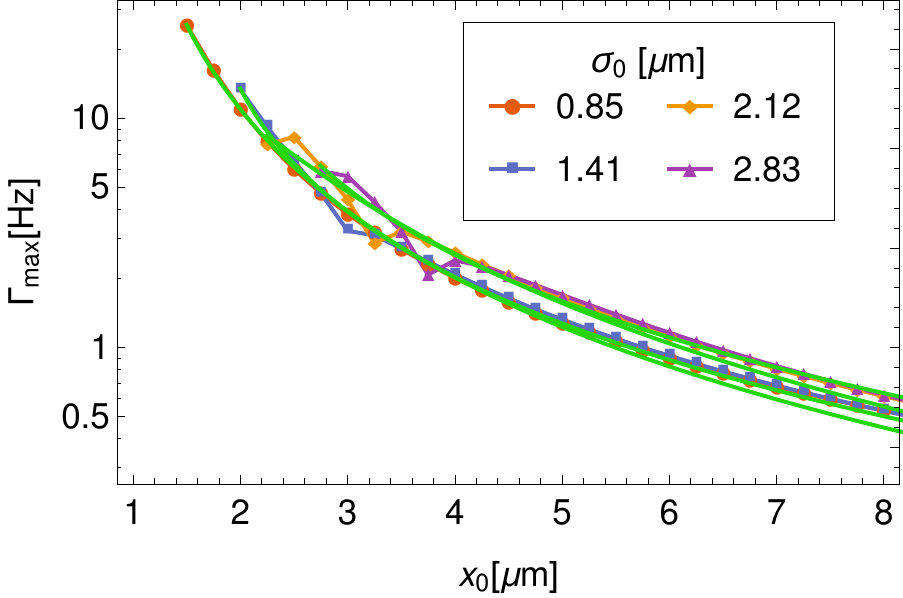}
	}
		\hfill
	\subfloat[]{
		\includegraphics[width=1.0\linewidth]{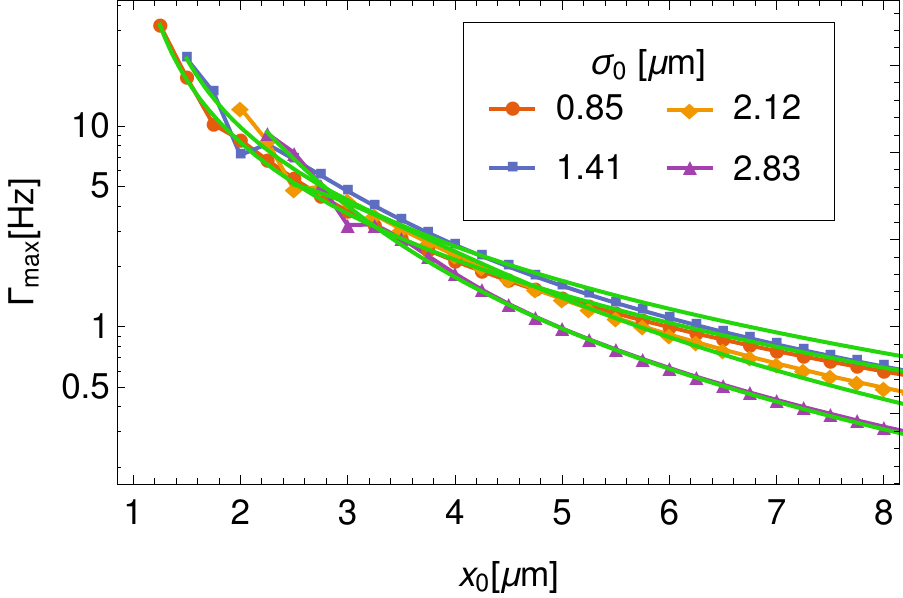}
	}
	\caption{\label{fig::sympot_x0_scaling}
		\textit{Scaling in barrier separation for symmetric trap.} Maximal tunneling rate $\Gamma_{\mathrm{max}}$ on a semi-log scale as a function of barrier separation for (a) $V_{0}=27.9 \mathrm{[nK]}$, and (b) $V_{0}=85.8 \mathrm{[nK]}$, with various Gaussian widths $\sigma_{0}$; solid green curves are fits and points are data (data points are connected as a guide to the eye).
		Tunneling rate decreases with both increasing $x_{0}$ and increasing $V_{0}$. All data exhibit a kink due to the mean-field islands for smaller $x_{0}$, but fits still capture the overall trend.  Larger $\sigma_{0}$ don't produce bound states for smaller traps. Error bars for data points are smaller than markers.
	}
\end{figure}

First, we examine scaling in the barrier separation $x_{0}$. Because the barrier height $V_{0}$ and width $\sigma_{0}$ are held constant, changes in the tunneling rate are largely due to an increase in the classical oscillation frequency, Eq.~\eqref{eq:oscillation}, which is strongly dependent to the separation $x_{0}$; in other words the exponential term in the tunneling rate, Eq.~\eqref{eq:gamma}, does not change much between adjacent $x_{0}$ for larger barrier separations when compared to the increase in oscillation frequency.
We plot and fit to $\Gamma_{\mathrm{max}}$ in Fig.~\ref{fig::sympot_x0_scaling} for several values of $\sigma_{0}$. For $\sigma_{0}=0.42$ an apparent \emph{kink} can be seen. This is due to the brief region of increasing $\Gamma_{\mathrm{max}}$ due to the appearance of mean-field islands, but is a weak effect for $x_{0}$ scaling.
Because the kink is less drastic than for $V_{0}$ scaling, we are able to fit a single curve to the entire domain. The data trends as
\begin{equation}
\Gamma_{\mathrm{max}} \simeq (a_{0}+a_{1} x_{0}+a_{2} x_{0}^{2})^{-1}, \label{eq:sym:x0_scaling}
\end{equation}
with fit parameters $a_{0}$, $a_{1}$, and $a_{2}$; this is a very rough trend due to the large deviation in $\Gamma_{\mathrm{max}}$ for smaller $x_{0}$ values.

Next, we examine scaling in $V_{0}$, which produces very different results. For $x_{0}=1.0,5.0 [\mathrm{\mu m}]$, we plot $\Gamma_{\mathrm{max}}$ vs $V_{0}$ for several values of $\sigma_{0}$ in Fig.~\ref{fig::sympot_v1_scaling}; the upper range of $\sigma_{0}$ values is approximately half of $x_{0}$ to allow sufficient room inside the trap. The effect due to the appearance of mean-field islands can be seen in Fig.~\ref{fig::sympot_v1_scaling}(a), with a dip followed by a steady rise at $V_{0} \approx 300 [\mathrm{nK}]$ for $\sigma_{0}=0.42 [\mathrm{\mu m}]$; smaller $\sigma_{0}$ barriers undergo this for a $V_{0}$ range too large to be captured in these plots. Barrier width has a large effect on the tunneling rates, with wider barriers reducing tunneling rates, as seen in Fig.~\ref{fig::sympot_v1_scaling}(a) and (b). Appearance of the mean-field islands causes a drastic change in $\Gamma_{\mathrm{max}}$, and so we use different fit functions before and after the kink. By examining the modified JWKB tunneling rate, Eq.~\eqref{eq:gamma}, to lowest order, we expect the tunneling rate to scale as $\Gamma_{\mathrm{max}} \simeq f(V_{0}) \exp[g(V_{0})]$, with suitable functions $f$ and $g$.

Before the kink, we find that the classical oscillation period dominates scaling, and so the data trends as
\begin{equation}
\Gamma_{\mathrm{max}} \simeq V_{0} (a_{0} + a_{1} V_{0}^{a_{2}})^{-1},
\end{equation}
as plotted in Fig.~\ref{fig::sympot_v1_scaling}(a); although we only plot data for $x_{0}=1.0$, we find for $x_{0}=1.0, 2.0, 3.0, 4.0, 5.0$ fit parameter $a_{2} = 1.32 \pm 0.05, 1.24 \pm 0.05, 1.23 \pm 0.03, 1.32 \pm 0.09, 2.0 \pm 0.8$ respectively. The factor of $V_{0}$ in the fit function likely comes from a weak linear dependence on the tunneling probability from Eq.~\eqref{eq:gamma}, and the power $a_{2}$ comes from the nonlinear dependence on $V_{0}$ in Eq.~\eqref{eq:momentum}. Although we can get arbitrarily small error in $\Gamma_{\mathrm{max}}$, assuming a $1\%$ error, we typically find a reduced chi-squared of $\chi_{r}^{2} \in [1,5]$. After the appearance of the effective mean-field islands, the penetration probability becomes important, and we find
\begin{equation}
\Gamma_{\mathrm{max}} \simeq a_{0} \sqrt{V_{0}} \exp{[-a_{1} V_{0}]}, 
\end{equation}
Fig.~\ref{fig::sympot_v1_scaling}(b); again assuming a $1\%$ uncertainty, we typically find $\chi_{r}^{2} \in [0.3,8.4]$.

\subsection{Scaling Laws for Tilt Barrier}

\begin{figure}
	\subfloat[]{
		\includegraphics[width=1.0\linewidth]{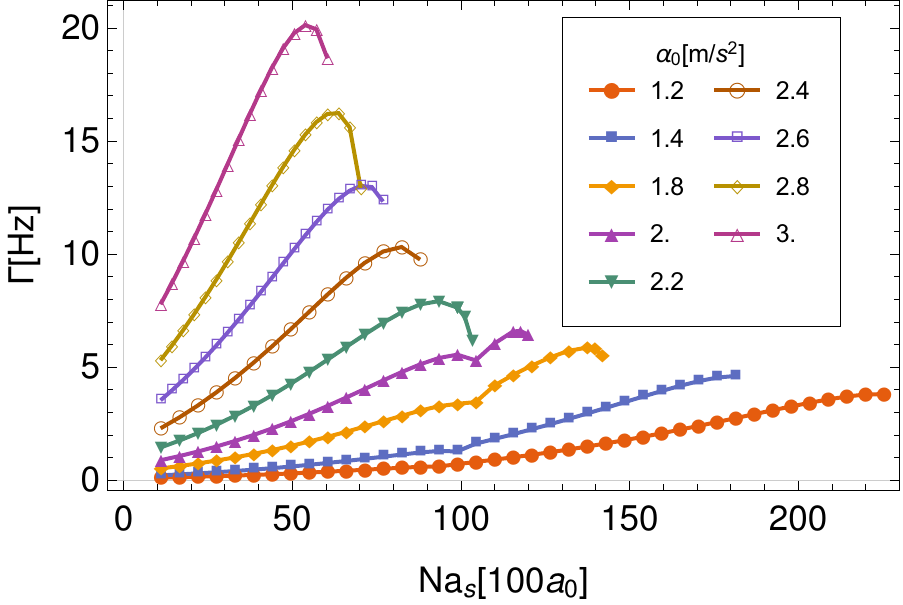}}
	\hfill
	\subfloat[]{
		\includegraphics[width=1.0\linewidth]{{{tilt_alpha_fits_sigma_0.6}}}
	}
	\caption{\label{fig::tilt_alpha_scaling}
		\textit{Scaling in ramp slope for tilt trap.}
		(a) Tunneling rate, $\Gamma$, is plotted as a function of interaction, in terms of effective scattering length $a_{s}$ in units of Bohr radius $a_{0}$ and number of atoms $N$, for several tilt $\alpha_{0}$, with $V_{0} = 79.3 \mathrm{[nK]}$ and $\omega_{\bot} = 2 \pi \times 1000 \mathrm{[Hz]}$. Emergence of effective mean-field islands is apparent for $\alpha_{0} = 2$ with appearance of a \emph{kink}. Markers are data points and curves are a guide to the eye.
		(b) Maximal tunneling rate $\Gamma_{\mathrm{max}}$ as a function of $\alpha_{0}$ for several barrier values. Fit curves are split with (solid green) and without (solid yellow) emergence of the effective mean-field islands. The largest barrier, $V_{0} = 125.6 \mathrm{[nK]}$, is sufficiently large to have mean-field islands for all plotted $\alpha_{0}$. All data points have error bars smaller than markers.
	}
\end{figure}

For $V_{\mathrm{tilt}}(x)$, the adjustable experimental parameters are the Gaussian height, $V_{0}$, Gaussian width, $\sigma_{0}$, acceleration, $\alpha_{0}$, and interaction, $Ng$. Similar to scaling in $V_{\mathrm{sym}}(x)$, $\sigma_{0}$ increases or decreases the overall trend in $\Gamma$, so we focus on scaling in $V_{0}$, $\alpha_{0}$, and $Ng$.

\begin{figure}
	\subfloat[]{
		\includegraphics[width=1.0\linewidth]{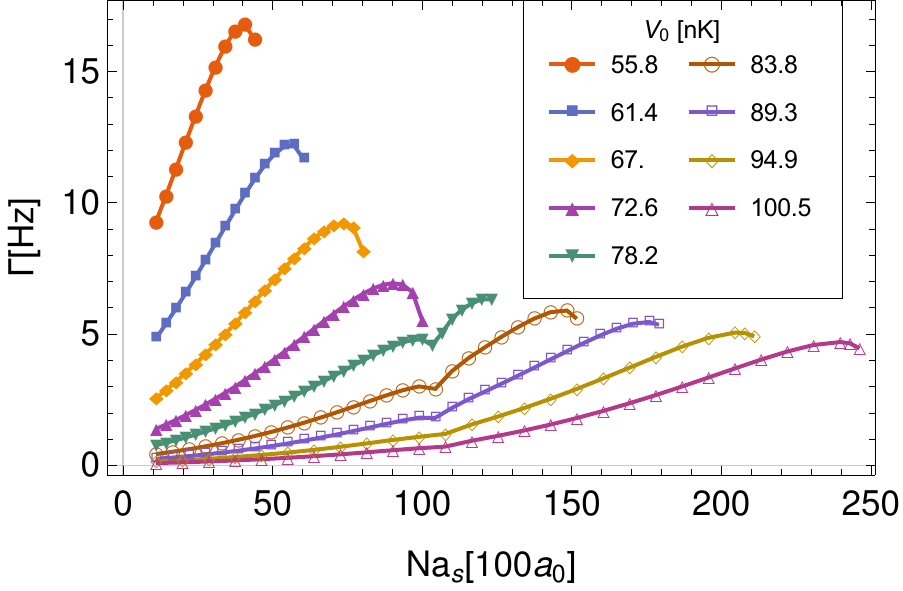}}
	\hfill
	\subfloat[]{
		\includegraphics[width=1.0\linewidth]{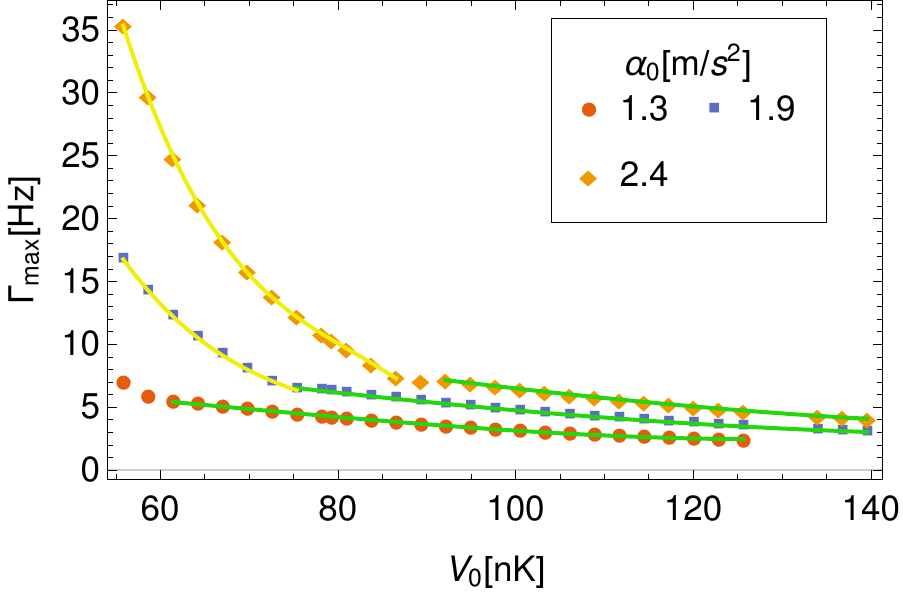}
	}
	\caption{\label{fig::tilt_v0_scaling}
		\textit{Scaling in barrier height for tilt trap.}
		(a) Tunneling rate, $\Gamma$, is plotted as a function of interaction, in terms of number of atoms $N$ and effective scattering length $a_{s}$ in units of Bohr radius $a_{0}$, for several barrier heights $V_{0}$, with $\alpha_{0} = 1.87 \mathrm{[m/s^{2}]}$ and $\omega_{\bot} = 2 \pi \times 1000 \mathrm{[Hz]}$. A kink in $\Gamma$ for $V_{0} = 78.2$ marks the appearance of effective mean-field islands. Markers are data, while curves are a guide to the eye.
		(b) Maximal tunneling rate, $\Gamma_{\mathrm{max}}$, as a function of $V_{0}$ for various $\alpha_{0}$ show clear trends before and after appearance of effective mean-field islands; fit curves are split before (solid yellow) and after (solid green) emergence of islands. All data points have error bars smaller than markers.
	}
\end{figure}

To qualitatively understand the role of $Ng$ in tunneling, we plot $\Gamma$ as a function of $N g$, in units of $s$-wave scattering $a_{s}$, for various barrier tilts, Fig.~\ref{fig::tilt_alpha_scaling}(a), and barrier heights, Fig.~\ref{fig::tilt_v0_scaling}(a).
The emergence of mean-field islands is very noticeable for $V_{\mathrm{tilt}}$. For $V_{\mathrm{sym}}(x)$ the emergence of mean-field islands for any given $\Gamma$ vs $g$ was not clear, only recognizable as a \emph{kink} in $\Gamma_{\mathrm{max}}$, while the effect is immediately noticeable for $\alpha_{0}=2 \mathrm{[m/s^{2}]}$ in Fig.~\ref{fig::tilt_alpha_scaling}(a) and $V_{0}=72.6 \mathrm{[nK]}$ in Fig.~\ref{fig::tilt_v0_scaling}(a). Similar to the symmetric trap, there exists a maximal interaction, $g_\mathrm{max}$, for any given barrier configuration where the repulsive interaction is sufficiently large to overcome the trapping potential, and no metastable bound states are possible.

Tunneling in $V_{\mathrm{tilt}}$ has a strong dependence on both $\alpha_{0}$ and $V_{0}$, in the sense that a change in either variable will necessarily result in simultaneously a wider or narrower trap and a smaller or larger trapping height for the atoms. For this reason, we are not able to describe scaling in any given variable directly to the classical oscillation period or tunneling probability, Eq.~\eqref{eq:gamma}.
Furthermore, unlike $V_{\mathrm{sym}}$, tunneling rate is not as strongly peaked about $N g$ for $V_{\mathrm{sym}}$; a change in $N g$ of 10\% can decrease $\Gamma$ by an order of magnitude or more in the symmetric trap. 
Although we can get arbitrarily small uncertainty in numerical data, all data is fit with an uncertainty of 1\%.
Scaling in $\alpha_{0}$ results in 2 distinct regimes, smaller (larger) alpha being with (without) the mean-field islands, depending on $V_{0}$.
For $\alpha_{0}$ with mean-field islands,  
\begin{equation}
\Gamma_{\mathrm{max}} \simeq a_{0} + a_{1} \alpha_{0}, \label{eq:tilt:alpha_scaling_1}
\end{equation}
with $\chi_{r}^{2} = 0.03$ and $\chi_{r}^{2} = 1.57$ for  $V_{0}=55.8$ and $79.3 \mathrm{[nK]}$ respectively.
The larger value of $V_{0}=125.6 \mathrm{[nK]}$ had
\begin{equation}
\Gamma_{\mathrm{max}} \simeq a_{0} + a_{2} \alpha_{0}^{2} + a_{3} \alpha_{0}^{3}, \label{eq:tilt:alpha_scaling_2}
\end{equation}
with $\chi_{r}^{2} = 0.69$; the much larger barrier has mean-field islands for the entire $\alpha_{0}$ range considered.
For $\alpha_{0}$ without mean-field islands, all data had fits of the form Eq.~\eqref{eq:tilt:alpha_scaling_2}; $V_{0}=55.8, 79.3 \mathrm{[nK]}$ with  $\chi_{r}^{2} = 0.98, 0.34$ respectively

Scaling in $V_{0}$ also gives 2 distinct regimes, again with and without mean-field islands. For smaller $V_{0}$, we find
\begin{equation}
\Gamma_{\mathrm{max}} \simeq a_{0} + a_{1} V_{0} + a_{2} V_{0}^{2} + a_{3} V_{0}^{3}, \label{eq:tilt:v0_scaling}
\end{equation} for $\alpha_{0} = 1.9$ and $2.4 \mathrm{[m/s^{2}]}$ with $\chi_{r}^{2} = 1.69, 0.71$ respectively. In the mean-field island regime, we find the smaller $\alpha_{0} = 1.3 \mathrm{[m/s^{2}]}$ resulted in scaling as Eq.~\eqref{eq:tilt:v0_scaling} except without the quadratic portion, $a_{2}=0$,
$\chi_{r}^{2} = 0.86$ with $\chi_{r}^{2} = 0.86$. 
The larger values of $\alpha_{0}= 1.9$ and $2.4 \mathrm{[m/s^{2}]}$ scale as
Eq.~\eqref{eq:tilt:v0_scaling} without the linear part, $a_{1}=0$,
with $\chi_{r}^{2} = 0.82$ and $0.05$ respectively.

\subsection{Experimental Implementation}
\label{sec:Experimental_Conditions}

A combination of magnetic and optical trapping techniques can be used to study macroscopic quantum escape in the two barrier configurations~\cite{weiner_experiments_1999}. Here we discuss the regimes which can be accessed experimentally and possible limitations. There are four major constraints that must be met: the chemical potential, $\mu$, mean field interaction, $Ng$, BEC number density, $n$, and tunneling time, $1/\Gamma$, must all be sufficiently small. Furthermore, we examine how the choice atomic species change these requirements. 

In order to keep the transverse wave function component in the ground state, the chemical potential in the quasi-1D trap must be below the traverse energy spacing, $\mu < \hbar \omega_{\bot}$. The chemical potential is limited by the trapping barrier height, $V_{0}$, giving an overall maximal barrier height $V_{\mathrm{max}}= \hbar \omega_{\bot}$, which ranges from $2.2 \mathrm{[nK]}$ for $\omega_{\bot} = 2 \pi \times 50 \mathrm{[Hz]}$ to $48 [\mathrm{nK}]$ for $\omega_{\bot}=2 \pi \times 1000 \mathrm{[Hz]}$.
For $V_{0}$ scaling in $V_{\mathrm{sym}}$, smaller potentials like $x_{0}= 1.0 \mathrm{\mu m}$, Fig.~\ref{fig::sympot_v1_scaling}(a), will be completely dominated by the classical oscillation period with short tunneling times $1/\Gamma = \mathcal{O}(0.01) \mathrm{[s]}$, and require tighter transverse confinement, $2 x_{0} \gg \ell_{\bot}$, i.e., $\ell_{\bot} \approx 0.4 \mathrm{[\mu m]}$ for $\omega_{\bot} = 2 \pi \times 500 \mathrm{[Hz]}$. Furthermore, for such small trapping well sizes, the Gaussian width must be sufficiently small, or a very shallow well will be created.
Larger wells, $x_{0} = 5.0 \mathrm{[\mu m]}$ with $\sigma_{0} = 1.7 \text{ to } 2.5 \mathrm{[\mu m]}$, as shown in Fig.~\ref{fig::sympot_x0_scaling}(b), allow access to regimes with and without mean-field islands; the transition occurs around $V_{0} = 15 \text{ to } 20 \mathrm{[nK]}$. 
Because $V_{\mathrm{tilt}}$ does allow direct control on well width, only indirectly by the ramp slope $\alpha_{0}$, we used Gaussian barriers with $\sigma_{0} = 0.85 \mathrm{[\mu m]}$, resulting in faster tunneling rates and requiring larger barrier heights $V_{0}$. 
In order to satisfy the bound $\mu < \hbar \omega_{\bot}$ for the regimes explored in this Article in $V_{\mathrm{tilt}}$, much tighter confinements, $\omega_{\bot}>1000 \mathrm{[Hz]}$, are required. This will cause  $n \geq \mathcal{O}(10^{15} \mathrm{[cm^{-3}]})$, causing 3-body loss to dominate for rubidium BECs.

The mean-field energy in the trap is limited by the chemical potential, which is bounded by the transverse energy, $\hbar \omega_{\bot}$.
We can find upper limits on $Ng^{1D}$, assuming the Thomas-Fermi approximation and replacing the potentials in Eq.~\eqref{eq:SymPot} and \eqref{eq:TiltPot} with infinite hard walls; $V_{\mathrm{sym}}$ becomes an infinite square well and $V_{\mathrm{tilt}}$ becomes an infinite square well with a linear ramp.
Using these assumptions in the GPE, one finds $N g_\mathrm{sym}^{1D} = 2 \mu x_{0}$ and $Ng_{\mathrm{tilt}}^{1D} = \frac{1}{2} \mu^{2} / \alpha_{0}$. 
These turn out to be good upper bounds when compared to numerical values of $Ng_{\mathrm{max}}^{1D}$ for any given potential configuration, which are typically $5 \text{ to } 10\%$ lower than the hard-wall Thomas-Fermi approximation. Using the bound $\mu < \hbar \omega_{\bot}$ and relation $g^{1D} = 2 \hbar \omega_{\bot} a_{s}$, we can derive limits for the number of particles and scattering length, $N a_{s}^{\mathrm{sym}} < \frac{1}{4} \hbar \omega_{\bot} / \alpha_{0}$ and $N a_{s}^{\mathrm{tilt}} <  x_{0}$. For rubidium in $V_{\mathrm{sym}}$ with unaltered scattering length $a_{s} \approx 98 a_{0}$, with Bohr radius $a_{0}$, one finds $N = 200 \text{ to } 1300$ for $x_{0} = 1 \text{ to } 7 \mathrm{[\mu m]}$, and tunneling times of $1/\Gamma = 0.05 \text{ to } 2.0 \mathrm{[s]}$. In contrast, $V_{\mathrm{sym}}$ with $\alpha_{0} =1 \mathrm{[m/s^{2}]}$, $\omega_{\bot} = 2 \pi \times 1000 \mathrm{[Hz]}$, and $a_{s} \approx 98 a_{0}$, a configuration which should allow larger number of atoms, gives only $N \approx 200$. Thus, $a_{s}$ would have to be decreased for rubidium to allow for several hundred atoms, on the order $a_{s} \lesssim \mathcal{O}(10)[a_{0}]$ for smaller $\omega_{\bot}$ and larger $\alpha_{0}$.

For the GPE, Eq.~\eqref{eq:gpe}, to be a proper quasi-1D description, the harmonic oscillator length, $\ell_{\bot} = \sqrt{\hbar / m \omega_{\bot}}$, must but be smaller than the healing length, $\xi = (8 \pi n a_{s})^{-1/2}$. This condition gives us an upper bound to the BEC density,
$n_{\mathrm{max}} = (8 \pi \ell_{\mathrm{\bot}}^{2} a_{s})^{-1}$.
Using $a_{s} = 1 \text{ to } 100 [a_{0}]$ and $\omega_{\bot} = 2 \pi \times (100 \text{ to } 1000 )\mathrm{[Hz]}$ give $n_{\mathrm{max}} = 10^{12} \text{ to } 10^{15} \mathrm{[cm^{-3}]}$, which is a density range most BEC experiments fall into. The larger densities, $10^{15} \mathrm{[cm]}$, will cause 3-body loss mechanics to become dominant on $\mathcal{O}(s)$~\cite{moerdijk_decay_1996}, and thus require sub-second tunneling times.

The condition on sufficiently fast tunneling rates, $\Gamma$, is arguably the most important since exceedingly long tunneling times are what prevent observable MQT in many systems, such as in chiral isomers, and need to be faster than 3-body loss rates~\cite{egorov_measurement_2013}.
Even with increased tunneling times for heavier objects, all figures in this Article present tunneling regimes with realizable rates using rubidium, one of the heavier atoms for BECs. 
Rubidium is often preferred for BEC experiments because the 3-body loss rate is two orders of magnitude smaller than lithium and sodium. However, since escape tunneling only requires the BEC to exist long enough to measure tunneling, we can overcome this limitation with faster escape rates.
All quantities in this Article can be easily rescaled for these lighter masses.
To be specific, for quantities such as $\mu$, $V_{0}$, $\alpha_{0}$, and $g$, rescaling can be determined from non-dimensionalization. For example, assuming some length scale $L$ and nondimenzionalized quantity, $\tilde{\mu}$, then
$\mu^{\mathrm{Li}} = \frac{\hbar^{2}}{m^{\mathrm{Li}} L^{2}} \tilde{\mu} = \frac{m^{\mathrm{Rb}}}{m^{\mathrm{Li}}} \mu^{\mathrm{Rb}}$.
In other words all gain a factor of $m^{Rb}/m^{\mathrm{Li}} \approx 12.5$, or for example $m^{\mathrm{Rb}}/m^{\mathrm{Na}} \approx 3.8$; this allows for larger traps, interaction strengths, and tunneling rates.
Rescaling $\Gamma$ is subtle, and depends on the exact question being asked. One possibility is,``given some specific values of $g$, $\mu$, and trap parameters, how does $\Gamma$ change for different atoms?" Going from rubidium to lithium gives $\Gamma^{\mathrm{Li}} = \Gamma^{\mathrm{Rb}} \sqrt{m^{\mathrm{Rb}}/m^{\mathrm{Li}}} P_{\mathrm{tunn}}^{\sqrt{m^{\mathrm{Li}}/m^{\mathrm{Rb}}} -1}$, where $P_{\mathrm{tunn}}= 0.05 \text{ to } 0.10$ are typical tunneling probabilities, giving $\Gamma^{\mathrm{Li}} \approx (10 \text{ to } 20 ) \Gamma^{\mathrm{Rb}}$. However, if instead we ask, ``how do we rescale $\Gamma_{\mathrm{max}}$ values from rubidium to other atoms, rescaling all parameter quantities $\mu$, $g$, etc?", the answer is simpler,
$ \Gamma^{\mathrm{Li}} = \frac{m^{\mathrm{Rb}}}{m^{\mathrm{Li}}} \Gamma^{\mathrm{Rb}}$.
Maximal density is rescaled as $n_{\mathrm{max}}^{Li}  = \frac{m^{\mathrm{Li}}}{m^{\mathrm{Rb}} } n_{\mathrm{max}}^{Rb}$. So, while we had maximal densities of $n_{\mathrm{max}}^{\mathrm{Rb}} = 10^{12 \text{ to } 15} \mathrm{[cm^{-3}]}$, we find $n_{\mathrm{max}}^{\mathrm{Li}} = 10^{11} \text{ to } 10^{14} \mathrm{[cm^{-3}]}$, which produces BECs that survive up to a second in the most extreme traps considered in this article. This is particularly important for $V_{\mathrm{tilt}}$, which requires tighter transverse confinement, and thus higher densities.

\section{Considerations on Variational JWKB}
\label{sec:Variational JWKB Considerations}

In this section we explore the emergence of small effective mean-field barriers which protrude above the chemical potential, appearing for sufficiently large $Ng$ and $V_{0}$. We also consider different wave function variational ans\"atze; the most commonly used pure Gaussian, to a linear ansatz with Gaussian tails appropriate to decay in a harmonic trap, to exponential tails more appropriate to decay in a square barrier.  Properly accounting for the mean field as well as making an informed choice of variational ansatz improves tunneling predictions enormously.

\subsection{Effective Potential: Single Well becomes Multiple-Well}
\label{sec:eff_mult_wells}

\begin{figure}
	\centering
	\includegraphics[width=1.0\linewidth]{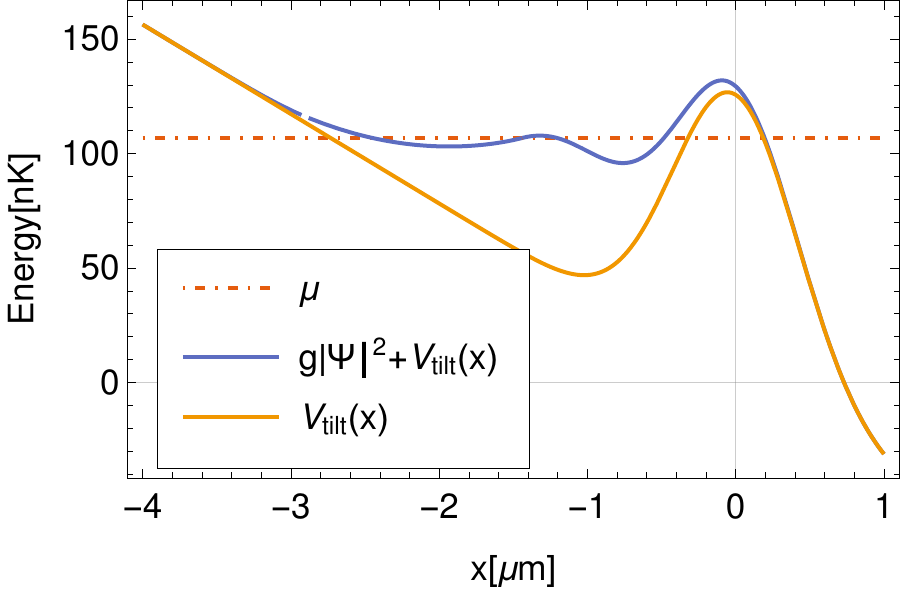}
	\caption{\label{fig::plot_mean-field-barrier} 
		\emph{Effective potential produces extra barrier.}
		The effective potential, $g | \Psi|^{2} + V_{\mathrm{tilt}}(x)$ (solid blue line), contains a small \emph{island} region, located around $x \approx -1.5$, which goes slightly above the chemical potential, $\mu$ (dash-dotted red line); the bare potential $V_{\mathrm{tilt}}(x)$ (solid yellow line) is shown for comparison. Potential parameters are $V_{0} = 125.6 \mathrm{[nK]}$, $\sigma_{0} =  0.85 \mathrm{[\mu m]}$, and $\alpha_{0} = 3.74 \mathrm{[m/s^{2}]}$. The symmetric potential, not shown, produces two islands.
	}
\end{figure}

In the well-known JWKB approximation, classical turning points are calculated by solving for zero momentum. This is modified in the variational-JWKB method, by using the effective mean-field potential in the semiclassical momentum, Eq.~\eqref{eq:momentum}. With the effective potential, we find that certain barrier parameters, with large enough nonlinearity, result in additional effective barriers with energies above the chemical potential, as shown in Fig.~\ref{fig::plot_mean-field-barrier}; although we only show the case for $V_{\mathrm{tilt}}$, we find similar results in the symmetric barrier, except in the symmetric case two barriers are produced inside the potential.
Consequently, the system is then composed of the trapping barriers and  multiple smaller barriers, or \emph{islands}.
Emergence of these multiple wells requires a slight modification when calculating the tunneling rates.
The probability to tunnel through these islands is symmetric; the probability of going from left to right is the same as from right to left. Along with this symmetric nature, the probability of tunneling through these islands is always much larger, by an order of magnitude, than the outermost barrier. 
We can then, on average, approximate the semi-classical period $T$ of one particle in the system as the sum of all periods, $T_i$, in every well $i$. The probability of passing through the outer-most barrier in one \emph{attack} is unaffected by these multiple wells. Hence, when these islands appear, we modify Eq.~\eqref{eq:gamma} to
\begin{equation}
\Gamma = \left(\sum\limits_{i} T_i\right)^{-1} \exp\left({-\frac{2}{\hbar}\int_{x_1}^{x_2} \mathrm{dx} |p(x)|}\right). \label{eq:gamma_mod}
\end{equation}

Island formation is robust, in the sense that islands will appear for sufficiently large interactions, $g'$, with appropriate barrier parameters, and then persist for all larger values $g>g'$.
Islands are caused by a combination of the self-interaction, $g$, and the wave function overlapping with a barrier, similar to how a water-wave \emph{swells} as it approaches a shoreline. However, while islands appear from both Gaussian barriers in $V_{\mathrm{sym}}$ overlapping with the wave function tails, for $V_{\mathrm{tilt}}$ only the left-hand tail (linear and Gaussian part) in Eq.~\eqref{eq:Tilt_WF} interacts with the linear potential to cause mean-field islands. As a consequence of this, mean-field islands will emerge for sufficiently small $\alpha'$, and then for all smaller $\alpha_{0} < \alpha'$, Fig.~\ref{fig::tilt_alpha_scaling}. Scaling in $x_{0}$ shows a similar situation, where $x_{0}$ must be sufficiently large for islands to form.

\subsection{Considerations on Variational Wave function: Tails and Connecting Functions}
\label{sec:Considerations_WF}

\begin{figure}
	\subfloat{
		\includegraphics[width=1.0\linewidth]{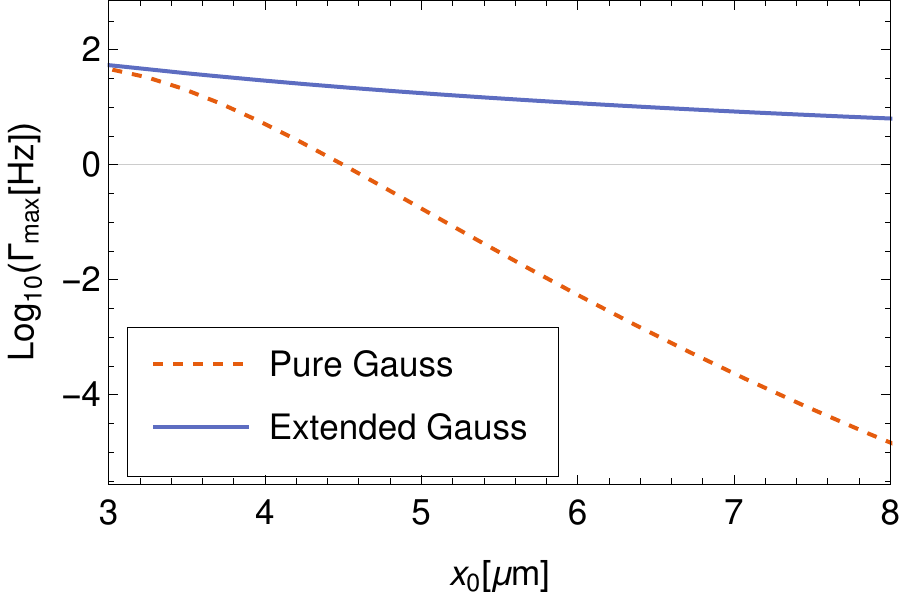}}
		\caption{\label{fig:compare_gaussian_to_flat_gaussian}
		\textit{Commonly used pure Gaussian variational ansatz severely underestimates tunneling rate.}
		Tunneling rate as a function of barrier separation, $x_{0}$, is plotted for a pure Gaussian (dashed red) and Gaussian with a flat middle region, extended Gaussian Eq.~\eqref{eq:Gauss_WF} (solid blue), in $V_{\mathrm{sym}}$ with parameters $V_{0} = 79.8 \mathrm{[nK]}$ and  $\sigma_{0} = 2.12 \mathrm{[\mu m]}$. The tunneling rates are almost equal for smaller separation, where the extended Gaussian approximates the pure Gaussian.
		}
\end{figure}

Gaussian wave functions are commonly used as a variational ansatz, from anomalous nonlinear ocean waves~\cite{ruban_gaussian_2015} to quantum droplets in BECs~\cite{wachtler_ground-state_2016}. However, in order to fully explore the parameter space for our trapping potentials, we extend a simple Gaussian by placing a linear function between the two tails in Eq.~\eqref{eq:Gauss_WF}, allowing the wave function to spread out in larger traps; this is especially important when the BEC is in the Thomas-Fermi operating regime. For larger traps, a pure Gaussian will stay largely localized inside the trap, which is unphysical and hence produce quantitatively wrong tunneling rates. To explicitly see this, we plot $\Gamma_{\mathrm{max}}$ for fixed $\sigma_{0}=1.5$  and $V_{0} = 79.8 \mathrm{[nK]}$ for increasing $x_{0}$ in Fig.~\ref{fig:compare_gaussian_to_flat_gaussian}. For sufficiently small barrier separation, the tunneling rates are close, but quickly diverge for larger $x_{0}$.

Beyond allowing the Gaussian tails to be separated, we also examine different wave function tails.
BECs in harmonic traps have Gaussian tails, while tunneling through barriers is often approximated by exponential tails, such as in the JWKB approximation, building on the exact solution for the square barrier. To explore the overlap between these two physical regimes, both of which are relevant to BECs, we also study a superposition of exponential and Gaussian wave functions in the symmetric potential. In other words, the simple Gaussian tails in Eq.~\eqref{eq:Gauss_WF} are replaced with the superposition of a Gaussian and an exponential. Because this ansatz wave function cannot be continuous in the first derivative, as the reader may verify for themselves, we can introduce a \emph{connecting function}, which smoothly connects the linear and tail portions of the wave function. 
We find that, whether using a connecting function to smooth the first derivative, or a simple linear function to allow the tails to extend outwards, the same result was always obtained, such that the contribution to overall tunneling rate was nearly zero from the connecting function's section; this is due to the connecting region always being significantly smaller than the trapping potentials, and to greatly simplify numerics one can just allow the Gaussian-exponential tails to connect directly to the flat region. 
The Gaussian-exponential tail ansatz was found to reduce the maximal mean-field interaction, $Ng_{\mathrm{max}}$, by at least $20\%$, such that, for example, the curves in Fig.~\ref{fig::gamma_vs_g_various} end for smaller interactions. Furthermore, the tunneling rate and chemical potential only changed by upwards of $5\%$ for any given interaction strength. All of these results combine to produce maximal tunneling rates one to two orders of magnitude smaller for the Gaussian-exponential ansatz.

\section{Conclusions}
\label{sec:Conclusion}

In summary, we have used a combination of the variational technique and a modified JWKB method to calculate macroscopic quantum escape rates of repulsive BECs described by the GPE, in two experimentally realizable quasi-1D potentials. For the variational ansatz, we used Gaussian tails separated by a linear connecting function, which approximate the Thomas-Fermi regime in larger traps. The two traps were a \emph{symmetric} trap created by two offset Gaussians, Fig.~\ref{fig::SymPot}, and a \emph{tilt} trap formed by a Gaussian and linear ramp, Fig.~\ref{fig::TiltPot}. Using the variational-JWKB method, which includes the nonlinear atom-atom interactions, we explored the parameter space created by mean-field strength and the potential-specific \emph{control knobs}, Gaussian height, separation, width, and ramp acceleration. Furthermore, we found substantially different scaling laws in the maximal tunneling rate, the rate with largest nonlinearity allowed in a potential configuration, from polynomial, to rational, to exponential in the various potential parameters.

We examined the assumptions under which the parameter ranges could be experimentally realized, carefully considering BEC number density, chemical potential, mean-field interaction strength, and tunneling times.
Although tunneling rates are in general hindered for large barriers and large masses, we found rates ranging from $\mathcal{O}(0.1)$ to $\mathcal{O}(100)$ hertz, with trapping-well dimension $\mathcal{O}(1)$ microns, Gaussian barrier widths from $0.1$ to $3$ microns, and $s$-wave scattering lengths from $1$ to $100$ Bohr radii for rubidium atoms with densities $\mathcal{O}(10^{12} \text{ to } 10^{15}) \mathrm{[cm^{-3}]}$. We found that lighter atoms, such as lithium, allow for tunneling rates, nonlinearities, barrier heights, and linear ramps an order of magnitude larger, with densities still allowing the BEC to survive several seconds. Lithium was also found to be more suitable for the tilt trap, as the relatively small well behind the Gaussian requires tight transverse trapping frequencies above $2 \pi \times 1000 \mathrm{[Hz]}$.
Estimates for the number of allowable trapped atoms depend strongly on the trapping configuration. The symmetric trap can support $N = 200 \text{ to } 1300$ atoms with barrier separations $1 \text{ to } 7 \mathrm{[\mu m]}$ for rubidium with an unaltered $s$-wave scattering length, allowing large quantities by reducing the scattering length. The tilt trap requires smaller scattering lengths for rubidium, below $10$ Bohr radii, and tighter transverse confinement, required by the smaller well size, limiting the BEC density. This makes lithium a more suitable choice for the tilt trap, even though the 3-body loss rates are larger, because the density will be an order of magnitude smaller than for rubidium.

Both traps were found to produce regions in trap parameters where additional effective mean-field barriers, \emph{islands}, were produced inside of the traps, Fig.~\ref{fig::plot_mean-field-barrier}. These islands altered scaling in maximal tunneling rates. Most notably, for the symmetric potential, we found that the classical oscillation frequency dominates scaling in barrier separation for small Gaussian heights, where mean-field islands do not form. For larger heights, in the regime where mean-field islands exist, we found that both the oscillation period and tunneling probability contribute, producing significantly different scaling. 
The maximal tunneling rate for the tilt potential was found to scale from linear to 3rd order polynomials for barrier height and ramp acceleration. 
Finally, we explored the effect of different variational wave function ans\"atze, finding that a pure Gaussian underpredicts tunneling rates for wide traps, and a Gaussian-exponential superposition predicts tunneling rates an order of magnitude smaller than the Gaussian tails connected by an intermediate linear region.

The findings in this Article can be used as a road map towards the realization of quasi-1D macroscopic quantum escape experiments in ultracold atom systems. These systems can also be tuned to lower particle numbers, where many-body effects can become important. Simulations taking many-body effects into account predict that trapped atoms will remain coherent but become incoherent as they escape, that the decay process will be exponential in time~\cite{lode_how_2012}, and that recently-escaped particles will influence the tunneling process~\cite{alcala_entangled_2017}. Coherence of the trapped atoms suggests that our mean-field analysis could have some extension into the many-body regime. In contrast to non-interacting predictions, the mean-field decay process has been predicted and measured as non-exponential in time~\cite{potnis_interaction-assisted_2017,schlagheck_nonexponential_2007}. A current experiment did not have the necessary resolution to separate out mean-field and possible many-body effects~\cite{zhao_macroscopic_2017}, and these quasi-1D experiments are a possible avenue, i.e., performing interference measurement of the escaped atoms. 
Future theoretical analysis can use the parameter regimes in this Article to examine the full dynamics, including many-body effects, and determine if the mean-field islands are a persistent phenomena and have a measurable effect.
Confinement in the tilt trap could be made stronger, over $10$ kHz, and possibly lead towards studying the tunneling dynamics of strongly interacting systems like the Tonks-Girardeau gas~\cite{petrov_regimes_2000}; a complementary approach to the few-body analysis~\cite{del_campo_decay_2006}.

\section{Acknowledgments}        

This work was supported by the Alexander von Humboldt Foundation, the
Heidelberg Center for Quantum Dynamics and the Deutsche Forschungsgemeinschaft under Project No. WE2661/11-1 and Collaborative Research Centre SFB 1225 (ISOQUANT), NSF under grants DMR-1407962, PHY-1520915, and OAC-1740130, and the AFOSR under grant FA9550-14-1-0287. This work was performed in part at the Aspen Center for Physics, which is supported by National Science Foundation grant PHY-1607611.
The calculations were carried out in part using the high performance computing resources provided by the Golden Energy Computing Organization at the Colorado School of Mines. We thank Rockson Chang, Kenji Maeda, Marc Repp, Aephraim Steinberg, and Xinxin Zhao for discussions.



%

\end{document}